\documentclass{article}
\usepackage{ace,amsmath,graphicx,url,times}
\usepackage{pgfplots}
\usepackage{tikz}
\usetikzlibrary{shapes,arrows}
\usetikzlibrary{fit,backgrounds}
\def\T60{{$\mathrm{T_{60}}$}}
\def\C50{{$\mathrm{C_{50}}$}}

\definecolor{mycolor1}{HTML}{D8D8D8}%
\definecolor{mycolor2}{HTML}{8181F7}%
\definecolor{mycolor3}{HTML}{FE9A2E}%

\newlength\figureheightx
\newlength\figurewidthx

\newlength\yyshift
\newlength\xyshift
\def\trainset{{\it trainSet}}
\def\devset{{\it devSet}}
\def\evalset{{\it evalSet}}

\title{Evaluating the Non-Intrusive Room Acoustics Algorithm with the ACE Challenge}

\name{{\em Pablo Peso Parada$^1$\sthanks{The research leading to these results has received funding from the European Union's Seventh Framework Programme (FP7/2007-2013) under grant agreement n$^\circ$ ITN-GA-2012-316969.}, Dushyant Sharma$^1$, Toon van Waterschoot$^2$, Patrick A. Naylor$^3$}}
\address{ $^1$ Voicemail-To-Text Research, Nuance Communications Inc., Marlow, UK\\ $^2$ Dept. of Electrical Engineering (ESAT-STADIUS/ETC), KU Leuven, Belgium\\ $^3$ Department of Electrical and Electronic Engineering, Imperial College London, UK\\ \texttt{\{pablo.peso, dushyant.sharma\}@nuance.com}\\ \texttt{toon.vanwaterschoot@esat.kuleuven.be,p.naylor@imperial.ac.uk}}

\begin{document}

\ninept
\maketitle

\begin{sloppy}

\begin{abstract}
We present a single channel data driven method for non-intrusive estimation of full-band reverberation time and full-band direct-to-reverberant ratio.
The method extracts a number of features from reverberant speech and builds a model using a recurrent neural network to estimate the reverberant acoustic parameters.
We explore three configurations	by including different data and also by combining the recurrent neural network estimates using a support vector machine.
Our best method to estimate DRR provides a Root Mean Square Deviation (RMSD) of 3.84 dB and a RMSD of 43.19 \% for \T60 estimation.
\end{abstract}


\section{Introduction}
\label{sec:intro}

Sound propagation from the source to the receiver placed in a room may follow multiple paths due to reflections from walls or objects in the enclosed space. This multipath propagation creates a reverberant sound which depends on the characteristics of the room and positions of both source and receiver. 
The reverberation time (\T60) characterizes the acoustic properties of an enclosed space and it is theoretically independent of the source-receiver distance.
Alternative objective measurements such as Direct-to-Reverberation Ratio (DRR) or clarity index (\C50) \cite{Naylor2010} may be employed to take into account this dimension.
The calculation of these measures of reverberation require an estimation of the Room Impulse Response (RIR), however in many real situations this information remains unavailable and these measures need to be non-intrusively estimated from the reverberant signal. 

Several methods have been proposed to blindly estimate \T60.
The method proposed by L{\"o}llmann et al. \cite{Lollmann2010} estimates the decay rate from a statistical model of the sound decay using the {\it maximum likelihood} (ML) approach and then from this decay rate the method finds the ML estimate for \T60. 
The Eaton et al. \cite{Eaton2013} \T60 estimator is based on spectral decay distributions. In this case the signal is filtered with uniform Mel-spaced filters and from the output of this filter bank the decay rate is computed by applying a least-square linear fit to the time-frequency log magnitude bins. The variance of the negative gradients in the distribution of decay rates is then mapped to \T60 with a polynomial function.
Falk and Chan \cite{Falk2010} proposed a method to compute the reverberation time in the modulation domain. The algorithm is based on the idea that low modulation frequency energy (below 20Hz) is barely affected by the reverberation level whilst high modulation frequency energy increases with the reverberation level. The estimator is created with a Support Vector Regressor (SVR) and with the ratio of the average of low modulation frequency energy to different averages of high modulation frequency energy as the input features. In addition, the overall ratio can be mapped to estimate directly the DRR parameter.
Kendrick et al. \cite{Kendrick2008} compare two methods to estimate from speech and music signals different room acoustic parameters, mainly \T60 and $\mathrm{C_{80}}$. The first one uses an artificial neural network with 40 features extracted by sampling the power spectrum density estimation of the sum of the Hilbert envelopes computed for certain frequency bands. The second method finds the cleanest sections of free decays in the signal to estimate with ML approach the decay curve and average this estimation to obtain the final estimator.
Although room acoustic parameters can be also estimated from multichannel recordings, such as \T60 \cite{Dumortier2014} or DRR \cite{Georganti2014}, or per frequency bin \cite{Doire2015}, this paper focuses on the problem of single-channel full-band room acoustic parameter estimation.

These measures of reverberation have been applied to estimate the perceived quality \cite{Vallado2013} or intelligibility \cite{Kuttruff2009} of reverberant recordings. These were also shown to predict speech recognition performance \cite{Fukumori2010} \cite{Sehr2010} \cite{Tsilfidis2013} \cite{Peso2014}. In addition to these applications, several de-reverberation algorithms use measures of reverberation to suppress reverberation in speech \cite{Naylor2010} \cite{Tsilfidis2013}\cite{Kawahara2012} \cite{Couvreur2001} \cite{Mohammed2012}, and so it is important 	to develop methods that estimate these measures directly from the reverberant signal.

We propose a non-intrusive (NIRA) method to estimate the room acoustic parameters based on extracting a number of per-frame features from the reverberant speech. A recurrent neural network is then employed to model the relationship between these features and the room acoustics parameters, i.e. DRR and \T60. This technique was tested on the single-channel configuration of the ACE challenge \cite{Eaton2015a} organized by the IEEE Audio and Acoustic Signal Processing Technical Committee to compare different approaches to estimate DRR and \T60.

The remainder of the paper is organized as follows.
Section \ref{sec:II} describes the method proposed in this work. 
In Section \ref{sec:III} the metrics used to evaluate the methods are introduced and
results obtained on the ACE Challenge database are detailed in Section \ref{sec:IV}.
Finally, in Section \ref{sec:V} the conclusions of this contribution are drawn.

\section{NIRA method}
\label{sec:II}
The method shown in Fig. \ref{fig:NIRA_workflow} computes a set of frame-based features from the signal using a window size of 20~ms and a 50\% overlap. Non-speech frames are dropped out with a Voice Activity Detector (VAD) using P.56 method \cite{ITU_T_P56}. This estimator was originally proposed for estimating \C50 from 8kHz speech signals in \cite{Peso2015J1} and extended to 16kHz signals in \cite{Peso2015J2}. In this work we have employed the latter configuration which estimates 134 frame-based features from the reverberant signal:

\tikzstyle{block} = [draw, rectangle, 
    minimum height=3em, minimum width=1em]
\tikzstyle{sum} = [draw, fill=black, circle,inner sep=0pt]
\tikzstyle{input} = [coordinate]
\tikzstyle{output} = [coordinate]
\tikzstyle{pinstyle} = [pin edge={to-,thin,black}]

\begin{figure}
\centering
\resizebox{\linewidth}{!}{\begin{tikzpicture}[auto, node distance=2cm,>=latex']
	\tikzstyle{every node}=[font=\normalsize]
    \node [input, name=input] {};
    \node [block, right of=input, node distance=1.5cm] (Normalization) {Norm.};
    \node [block, right of=Normalization, node distance=1.2cm] (VAD) {VAD};
    \node [block, right of=VAD, node distance=1.6cm] (FrameBased) [align=center] {Frame-based\\ features};
    \node [block, right of=FrameBased, node distance=2.8cm] (LearningAlgorithm) [align=center] {Learning\\ algorithm};

    \node [output, right of=LearningAlgorithm] (output) {};

    \draw [draw,->] (input) -- node [align=center] {Speech\\ signal} (Normalization);
    \draw [draw,->] (Normalization) -- (VAD);
    \draw [->] (VAD) --  (FrameBased);
    \draw [->] (FrameBased) -- node {$\phi_{k,1:134}$} (LearningAlgorithm);
    \draw [->] (LearningAlgorithm) -- node [name=y] [align=center] {Room \\ acoustic \\ parameter \\ estimate}(output);
    \node[style={draw=gray,dashed,fill=none,thick,inner sep=5pt, label={[shift={(7ex,-3ex)}]north west:\textbf{NIRA}}},fit=(input) (Normalization) (VAD) (FrameBased) (LearningAlgorithm) (y)] (A) [align= left] {};
\end{tikzpicture}}
    	\caption{The NIRA method.}
	\label{fig:NIRA_workflow}
\end{figure}
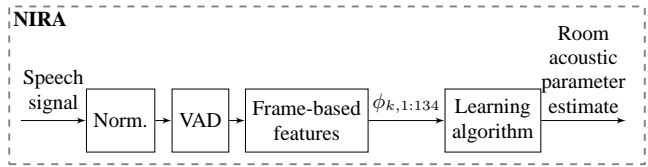

\begin{itemize}
	 	\item Line Spectrum Frequency (LSF) features computed by mapping the first 20 linear prediction coefficients to the LSF representation and their rate of change.
		\item Zero-crossing rate and its rate of change.
		\item Speech variance and its rate of change.
		\item Pitch period estimated with the PEFAC algorithm \cite{Gonzalez2011} and its rate of change.
		\item Estimation of the importance-weighted Signal-to-Noise Ratio (iSNR) in units of dB and its rate of change.
		\item Variance and dynamic range of the Hilbert envelope  and their rate of change.		
		\item Three parameters extracted from the Power spectrum of the Long term Deviation (PLD): spectral centroid, spectral dynamics and spectral flatness. The PLD is calculated per frame using the log difference between the signal power spectrum and long term average speech spectrum. Their rate of change is also included.
		\item 12th order mean- and variance-normalized Mel-frequency cepstral coefficients computed from the fast Fourier transform with delta and delta-delta.
		\item Modulation domain features \cite{Wang2014} derived from computing the first four central moments of the highest energy frequency band and its two adjacent modulation frequency bands.
		\item Deep scattering spectrum features are extracted from a scattering transformation applied to the signal \cite{Anden2014}.
\end{itemize}

These features are used to train a Bidirectional Long-Short Term Memory (BLSTM) \cite{Weninger2014a} recurrent neural network to provide an estimate of DRR and \T60 every 10 ms. The main motivation for using this architecture is that it can model temporal correlation such as reverberation due to its feedback connections. Alternative learning algorithms as classification and regression tree, linear regression or deep belief neural network have been investigated in the frame of \C50 estimation however BLSTM showed a better performance \cite{Peso2015J1}. Since ACE Challenge data assumes that the room acoustic properties remain unchanged within each utterance, only the temporal average for each utterance of all per frame estimations is considered.

Different architectures of the BLSTM\footnote{http://sourceforge.net/projects/currennt/} are explored with one to four layers including 64, 128 and 256 neurons per layer  and a minibatch size of 25, 50, 100 and 200 samples. 
Three different configurations were explored using this framework which are described in the following subsections.

\subsection{NIRAv1}
\label{subsec:NIRAv1}
This configuration is based on training the NIRA framework presented in Fig. \ref{fig:NIRA_workflow} using only the ACE Challenge development database. All data from the different microphone configurations was split randomly into three parts: training set ({\trainset}), development set ({\devset}) and evaluation set ({\evalset}). The {\trainset} comprises 70\% of the files in the ACE Challenge development database, whereas {\devset} and {\evalset} comprise 20~\% and 10~\% respectively. In this case, {\trainset} is used to train the model and {\devset} is employed to validate the model, then the selected model is the one that minimizes the estimation error in {\devset}.

\subsection{NIRAv2}
This configuration employs the NIRA framework shown in Fig. \ref{fig:NIRA_workflow} trained on three different databases in order to introduce new data in the model which could generalize the model to a wider range of scenarios. In this case 60\% of the files are extracted from the ACE Challenge development database, 20\% of the files from the REVERB Challenge database and the remainder of the files are taken from a database created with TIMIT database \cite{Garofolo1988} and real impulse responses from MARDY \cite{Wen2006}, SMARD \cite{Nielsen2014}, C4DM RIR \cite{Stewart2010} and REVERB Challenge \cite{REVERBchall} database. Similarly, {\devset} is created with the same proportions and from the same databases but the total number of files is 30\% of {\trainset}.

\subsection{NIRAv3}
This configuration follows the structure shown in Fig. \ref{fig:NIRAV3}. It is based on training 4 different BLSTM models using different data: NIRAv1; NIRA$_\alpha$ using the whole REVERB Challenge development set; NIRA$_\beta$ and NIRA$_\gamma$ employing real and simulated RIRs respectively convolved with TIMIT database. The real RIRs are taken from  MARDY, SMARD, C4MD and REVERB Challenge database, while the simulated RIRs are created with the randomized image method \cite{Desena2015}. These 4 estimators are combined by averaging the per-frame estimations of each utterance and by training a SVR model \cite{LIBSVM} with the 4-dimensional estimate vector obtained from the individual estimators. The training data for this SVR is {\devset} from NIRAv1 and {\evalset} is used for validation purposes.

\tikzstyle{block} = [draw, rectangle] 
\tikzstyle{sum} = [draw, fill=black, circle,inner sep=0pt]
\tikzstyle{input} = [coordinate]
\tikzstyle{output} = [coordinate]
\tikzstyle{pinstyle} = [pin edge={to-,thin,black}]

\begin{figure}
\centering
\resizebox{0.8\linewidth}{!}{\begin{tikzpicture}[auto,  node distance=2cm,>=latex']
	\tikzstyle{every node}=[font=\normalsize]
    \node [input, name=input] {};
    \node [sum, right of=input, node distance=0.9cm] (sum) {};
    \node [right of=sum, minimum size=0cm, node distance=1cm] (dummy) {}; 
    	\node [sum, right of=dummy, node distance=2.4cm] (sum2) {};
    \node [block, right of=dummy, node distance=3.2cm] (LearningAlgorithm) [align=center] {SVR};

    \node [output, right of=LearningAlgorithm] (output) {};
    \node [block, above of=dummy,node distance=0.5cm] (NIRA2) [align=center] {NIRA$_\alpha$};
    \node [above of=NIRA2, minimum size=0cm, node distance=0.5cm] (dummyab) {}; 
    \node [block, above of=dummyab,node distance=0.5cm] (NIRA1) [align=center] {NIRAv1};
    \node [block, below of=dummy,node distance=0.5cm] (NIRA3) [align=center] {NIRA$_\beta$};
    \node [below of=NIRA3, minimum size=0cm, node distance=0.5cm] (dummybe) {}; 
    \node [block, below of=dummybe,node distance=0.5cm] (NIRA4) [align=center] {NIRA$_\gamma$};

    \node [block, right of=NIRA1,node distance=1.5cm] (avg1) [align=center] {Temporal\\ averg.};    
    \node [block, right of=NIRA2,node distance=1.5cm] (avg2) [align=center] {Temporal\\ averg.};
    \node [block, right of=NIRA3,node distance=1.5cm] (avg3) [align=center] {Temporal\\ averg.};
    \node [block, right of=NIRA4,node distance=1.5cm] (avg4) [align=center] {Temporal\\ averg.};

    \draw [draw,->] (input) -- node [align=center] {Speech\\ signal} (sum);
    \draw [->] (sum) |- node {} (NIRA1);
    \draw [->] (sum) |- node {} (NIRA2);
    \draw [->] (sum) |- node {} (NIRA3);
    \draw [->] (sum) |- node {} (NIRA4);

    \draw [-] (NIRA1) -- node {} (avg1);
    \draw [-] (NIRA2) -- node {} (avg2);
    \draw [-] (NIRA3) -- node {} (avg3);
    \draw [-] (NIRA4) -- node {} (avg4);

    \draw [-] (avg1) -| node {} (sum2);    
    \draw [-] (avg2) -| node {} (sum2);
    \draw [-] (avg3) -| node {} (sum2);    
    \draw [-] (avg4) -| node {} (sum2);    
    
    \draw [->] (sum2) -- (LearningAlgorithm);
    
    \draw [->] (LearningAlgorithm) -- node [name=y] [align=center] {Room \\ acoustic \\ parameter \\ estimate}(output); 

\end{tikzpicture}}
    	\caption{The NIRAv3 method for DRR and \T60 estimation.}
	\label{fig:NIRAV3}
\end{figure}
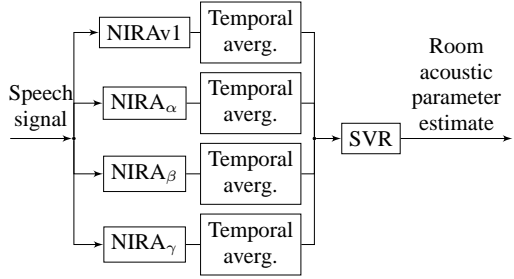

\section{Performance evaluation}
\label{sec:III}
All methods described in this paper are evaluated on the Acoustic Characterization of Environments (ACE) challenge \cite{Eaton2015a}. This challenge provides a common framework where different approaches of estimating DRR and \T60 can be directly compared. In addition to the box plots provided by the challenge to compare the different approaches, the algorithms are compared in this paper in terms of Root Mean Square Deviation (RMSD). This metric is computed for the DRR estimators as
\begin{equation}
\mathrm{RMSD_{DRR}}=\sqrt{\frac{\sum_{n=1}^{N} (\mathrm{\widehat{DRR_n}}-\mathrm{DRR_n})^2}{N}} \; \mathrm{dB},
\end{equation}
where $\mathrm{DRR_n}$ and $\mathrm{\widehat{DRR_n}}$ are the ground truth and the estimated DRR respectively of the $n$-th utterance  and $N$ is the total number of utterances.

On the other hand, the RMSD of the \T60 estimators is calculated as
\begin{equation}
\mathrm{RMSD_{T_{60}}}=\sqrt{\frac{\sum_{n=1}^{N} (100 \cdot ( \mathrm{\widehat{T_{60_n}}} - \mathrm{T_{60_n}})/\mathrm{T_{60_n}})^2}{N}} \; \mathrm{\%},
\end{equation}
where $\mathrm{T_{60_n}}$ and $\mathrm{\widehat{T_{60_n}}}$ are the ground truth and the estimated \T60 respectively.

\section{Results}
\label{sec:IV}
The evaluation results for the different approaches are shown in this section. These approaches are tested on two datasets: {\evalset} described in Section \ref{subsec:NIRAv1} and the ACE Challenge evaluation set.
\subsection{Performance in {\evalset}}
Table \ref{tab:RMSD_evalset} shows the performance of the three approaches in terms of RMSD on the {\evalset} dataset introduced in Section \ref{subsec:NIRAv1}. \mbox{NIRAv1} and  \mbox{NIRAv3} show the best performance for DRR estimation and NIRAv2 the highest estimation error deviation. Figure \ref{fig:BoxPlot_DRR_evalset} displays the box plot for the same dataset. NIRAv2 shows a wider interquartile range (IQR) and a negative bias which explains the higher RMSD value compared to the other two methods. Regarding \T60 estimation, Tab. \ref{tab:RMSD_evalset} indicates that the best approach is NIRAv1, whereas NIRAv3 provides the lowest performance mainly due to the bias and the wide IQR displayed in Fig. \ref{fig:BoxPlot_T60_evalset}.

\begin{table}[htb]
	\centering
	\begin{tabular}{r|cc} 
	{\bf Method} & {$\mathbf{RMSD_{DRR} \; (dB)}$} & {$\mathbf{RMSD_{T_{60}}\;(\%)}$}\\\hline
	NRIAv1 & 0.64 & 3.18\\
	NRIAv2 & 0.92 & 3.66\\
	NRIAv3 & 0.63 & 7.15\\
	\end{tabular}
	\caption{RMSD of the three approaches to estimate DRR and {\T60} using {\evalset} dataset.}
	\label{tab:RMSD_evalset}
\end{table}

\setlength\figureheightx{5cm}
\setlength\figurewidthx{0.8\linewidth}
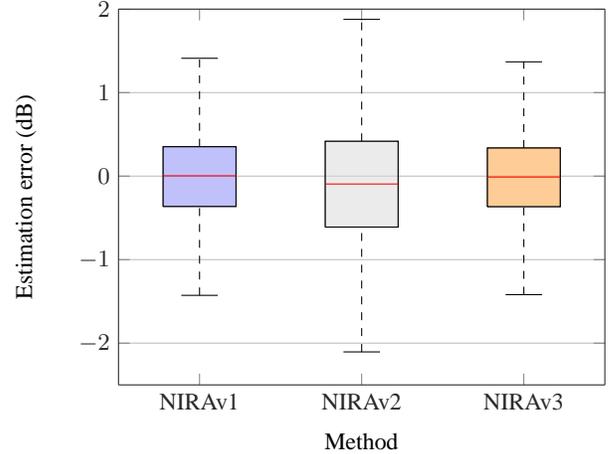
\begin{figure}[htc!]
\centering
%
%
%
%
\begin{tikzpicture}

\begin{axis}[%
width=0.950920245398773\figurewidthx,
height=\figureheightx,
at={(0\figurewidthx,0\figureheightx)},
scale only axis,
separate axis lines,
every outer x axis line/.append style={white!15!black},
every x tick label/.append style={font=\color{white!15!black}},
xmin=0.5,
xmax=3.5,
xtick={1,2,3},
xticklabels={{NIRAv1},{NIRAv2},{NIRAv3}},
xlabel={Method},
every outer y axis line/.append style={white!15!black},
every y tick label/.append style={font=\color{white!15!black}},
ylabel style={yshift=\yyshift},xlabel style={yshift=\xyshift},
ymin=-2.5,
ymax=2,
ylabel={Estimation error (dB)},
ymajorgrids,
yminorgrids
]

\addplot[area legend,solid,fill=mycolor3,opacity=5.000000e-01,draw=black,forget plot]
table[row sep=crcr] {%
x	y\\
2.775	-0.366314\\
2.775	0.3374125\\
3.225	0.3374125\\
3.225	-0.366314\\
2.775	-0.366314\\
}--cycle;

\addplot[area legend,solid,fill=mycolor1,opacity=5.000000e-01,draw=black,forget plot]
table[row sep=crcr] {%
x	y\\
1.775	-0.610110543720444\\
1.775	0.41760443910583\\
2.225	0.41760443910583\\
2.225	-0.610110543720444\\
1.775	-0.610110543720444\\
}--cycle;

\addplot[area legend,solid,fill=mycolor2,opacity=5.000000e-01,draw=black,forget plot]
table[row sep=crcr] {%
x	y\\
0.775	-0.363655384960299\\
0.775	0.352417486470387\\
1.225	0.352417486470387\\
1.225	-0.363655384960299\\
0.775	-0.363655384960299\\
}--cycle;

\addplot [color=black,dashed,forget plot]
  table[row sep=crcr]{%
1	0.352417486470387\\
1	1.41107859122402\\
};
\addplot [color=black,dashed,forget plot]
  table[row sep=crcr]{%
2	0.41760443910583\\
2	1.87690195041322\\
};
\addplot [color=black,dashed,forget plot]
  table[row sep=crcr]{%
3	0.3374125\\
3	1.368491\\
};
\addplot [color=black,dashed,forget plot]
  table[row sep=crcr]{%
1	-1.42760172007076\\
1	-0.363655384960299\\
};
\addplot [color=black,dashed,forget plot]
  table[row sep=crcr]{%
2	-2.10502212158809\\
2	-0.610110543720444\\
};
\addplot [color=black,dashed,forget plot]
  table[row sep=crcr]{%
3	-1.41932\\
3	-0.366314\\
};
\addplot [color=black,solid,forget plot]
  table[row sep=crcr]{%
0.8875	1.41107859122402\\
1.1125	1.41107859122402\\
};
\addplot [color=black,solid,forget plot]
  table[row sep=crcr]{%
1.8875	1.87690195041322\\
2.1125	1.87690195041322\\
};
\addplot [color=black,solid,forget plot]
  table[row sep=crcr]{%
2.8875	1.368491\\
3.1125	1.368491\\
};
\addplot [color=black,solid,forget plot]
  table[row sep=crcr]{%
0.8875	-1.42760172007076\\
1.1125	-1.42760172007076\\
};
\addplot [color=black,solid,forget plot]
  table[row sep=crcr]{%
1.8875	-2.10502212158809\\
2.1125	-2.10502212158809\\
};
\addplot [color=black,solid,forget plot]
  table[row sep=crcr]{%
2.8875	-1.41932\\
3.1125	-1.41932\\
};
\addplot [color=black,solid,forget plot]
  table[row sep=crcr]{%
0.775	-0.363655384960299\\
0.775	0.352417486470387\\
1.225	0.352417486470387\\
1.225	-0.363655384960299\\
0.775	-0.363655384960299\\
};
\addplot [color=black,solid,forget plot]
  table[row sep=crcr]{%
1.775	-0.610110543720444\\
1.775	0.41760443910583\\
2.225	0.41760443910583\\
2.225	-0.610110543720444\\
1.775	-0.610110543720444\\
};
\addplot [color=black,solid,forget plot]
  table[row sep=crcr]{%
2.775	-0.366314\\
2.775	0.3374125\\
3.225	0.3374125\\
3.225	-0.366314\\
2.775	-0.366314\\
};
\addplot [color=red,solid,forget plot]
  table[row sep=crcr]{%
0.775	0.00325215825242697\\
1.225	0.00325215825242697\\
};
\addplot [color=red,solid,forget plot]
  table[row sep=crcr]{%
1.775	-0.0944853768015683\\
2.225	-0.0944853768015683\\
};
\addplot [color=red,solid,forget plot]
  table[row sep=crcr]{%
2.775	-0.01078\\
3.225	-0.01078\\
};
\end{axis}
\end{tikzpicture}%
	\caption{Distribution of the DRR estimation errors for each method using {\evalset}. The edges of the boxes indicate the lower and upper quartile range, while the horizontal lines inside the boxes represent the medians for each method. Moreover, the horizontal lines outside the boxes indicate the estimation error up to 1.5 times the interquartile range.}
	\label{fig:BoxPlot_DRR_evalset}
\end{figure} 

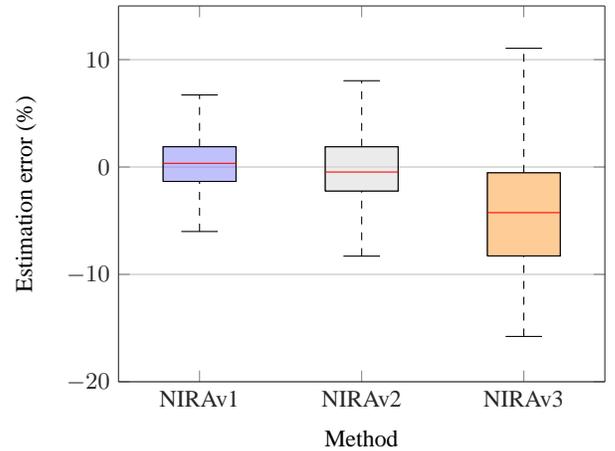
\begin{figure}[htc!]
\centering
%
%
%
%
\begin{tikzpicture}

\begin{axis}[%
width=0.950920245398773\figurewidthx,
height=\figureheightx,
at={(0\figurewidthx,0\figureheightx)},
scale only axis,
separate axis lines,
every outer x axis line/.append style={white!15!black},
every x tick label/.append style={font=\color{white!15!black}},
xmin=0.5,
xmax=3.5,
xtick={1,2,3},
xticklabels={{NIRAv1},{NIRAv2},{NIRAv3}},
xlabel={Method},
every outer y axis line/.append style={white!15!black},
every y tick label/.append style={font=\color{white!15!black}},
ylabel style={yshift=\yyshift},xlabel style={yshift=\xyshift},
ymin=-20,
ymax=15,
ylabel={Estimation error (\%)},
ymajorgrids,
yminorgrids
]

\addplot[area legend,solid,fill=mycolor3,opacity=5.000000e-01,draw=black,forget plot]
table[row sep=crcr] {%
x	y\\
2.775	-8.28258739032897\\
2.775	-0.534249890779471\\
3.225	-0.534249890779471\\
3.225	-8.28258739032897\\
2.775	-8.28258739032897\\
}--cycle;

\addplot[area legend,solid,fill=mycolor1,opacity=5.000000e-01,draw=black,forget plot]
table[row sep=crcr] {%
x	y\\
1.775	-2.24201302508113\\
1.775	1.88564927532387\\
2.225	1.88564927532387\\
2.225	-2.24201302508113\\
1.775	-2.24201302508113\\
}--cycle;

\addplot[area legend,solid,fill=mycolor2,opacity=5.000000e-01,draw=black,forget plot]
table[row sep=crcr] {%
x	y\\
0.775	-1.3431626572466\\
0.775	1.89140560058307\\
1.225	1.89140560058307\\
1.225	-1.3431626572466\\
0.775	-1.3431626572466\\
}--cycle;

\addplot [color=black,dashed,forget plot]
  table[row sep=crcr]{%
1	1.89140560058307\\
1	6.71353672136588\\
};
\addplot [color=black,dashed,forget plot]
  table[row sep=crcr]{%
2	1.88564927532387\\
2	8.03457722687204\\
};
\addplot [color=black,dashed,forget plot]
  table[row sep=crcr]{%
3	-0.534249890779471\\
3	11.0589242452479\\
};
\addplot [color=black,dashed,forget plot]
  table[row sep=crcr]{%
1	-6.00689303833667\\
1	-1.3431626572466\\
};
\addplot [color=black,dashed,forget plot]
  table[row sep=crcr]{%
2	-8.29762635381185\\
2	-2.24201302508113\\
};
\addplot [color=black,dashed,forget plot]
  table[row sep=crcr]{%
3	-15.7792547032116\\
3	-8.28258739032897\\
};
\addplot [color=black,solid,forget plot]
  table[row sep=crcr]{%
0.8875	6.71353672136588\\
1.1125	6.71353672136588\\
};
\addplot [color=black,solid,forget plot]
  table[row sep=crcr]{%
1.8875	8.03457722687204\\
2.1125	8.03457722687204\\
};
\addplot [color=black,solid,forget plot]
  table[row sep=crcr]{%
2.8875	11.0589242452479\\
3.1125	11.0589242452479\\
};
\addplot [color=black,solid,forget plot]
  table[row sep=crcr]{%
0.8875	-6.00689303833667\\
1.1125	-6.00689303833667\\
};
\addplot [color=black,solid,forget plot]
  table[row sep=crcr]{%
1.8875	-8.29762635381185\\
2.1125	-8.29762635381185\\
};
\addplot [color=black,solid,forget plot]
  table[row sep=crcr]{%
2.8875	-15.7792547032116\\
3.1125	-15.7792547032116\\
};
\addplot [color=black,solid,forget plot]
  table[row sep=crcr]{%
0.775	-1.3431626572466\\
0.775	1.89140560058307\\
1.225	1.89140560058307\\
1.225	-1.3431626572466\\
0.775	-1.3431626572466\\
};
\addplot [color=black,solid,forget plot]
  table[row sep=crcr]{%
1.775	-2.24201302508113\\
1.775	1.88564927532387\\
2.225	1.88564927532387\\
2.225	-2.24201302508113\\
1.775	-2.24201302508113\\
};
\addplot [color=black,solid,forget plot]
  table[row sep=crcr]{%
2.775	-8.28258739032897\\
2.775	-0.534249890779471\\
3.225	-0.534249890779471\\
3.225	-8.28258739032897\\
2.775	-8.28258739032897\\
};
\addplot [color=red,solid,forget plot]
  table[row sep=crcr]{%
0.775	0.343072460891586\\
1.225	0.343072460891586\\
};
\addplot [color=red,solid,forget plot]
  table[row sep=crcr]{%
1.775	-0.46808788221564\\
2.225	-0.46808788221564\\
};
\addplot [color=red,solid,forget plot]
  table[row sep=crcr]{%
2.775	-4.24965778792525\\
3.225	-4.24965778792525\\
};
\end{axis}
\end{tikzpicture}%
	\caption{Distribution of the \T60 estimation errors for each method using {\evalset}.}
	\label{fig:BoxPlot_T60_evalset}
\end{figure}

\subsection{Performance in ACE Challenge Evaluation set}
Table \ref{tab:RMSD_evalACE} shows the performance of the three approaches on the ACE Challenge evaluation dataset. NIRAv3 and NIRAv1 still provide the best performance when estimating DRR and \T60 respectively on this dataset, however the deviations are considerably increased.
\begin{table}[htb]
	\centering
	\begin{tabular}{r|cc} 
	{\bf Method} & {$\mathbf{RMSD_{DRR}\;(dB)}$} & {$\mathbf{RMSD_{T_{60}}\;(\%)}$}\\\hline
	NRIAv1 & 3.87 & 43.19 \\
	NRIAv2 & 3.85 & 44.80\\
	NRIAv3 & 3.84 & 44.18\\
	\end{tabular}
	\caption{RMSD of the three approaches to estimate DRR and {\T60} using ACE Challenge evaluation set.}
	\label{tab:RMSD_evalACE}
\end{table}

Figure \ref{fig:BoxPlot_DRR_ACE} shows the distribution of the DRR estimation error for each method. The three methods present similar distributions, however NIRAv3 is less biased which is in accordance with the results displayed in Tab. \ref{tab:RMSD_evalACE}. 
Figure \ref{fig:BoxPlot_T60_ACE} shows the box plot for each method proposed to estimate \T60. NIRAv3 presents the higher interquartile range and NIRAv1 the least biased estimation, which is reflected in the deviation shown in Tab. \ref{tab:RMSD_evalACE}.

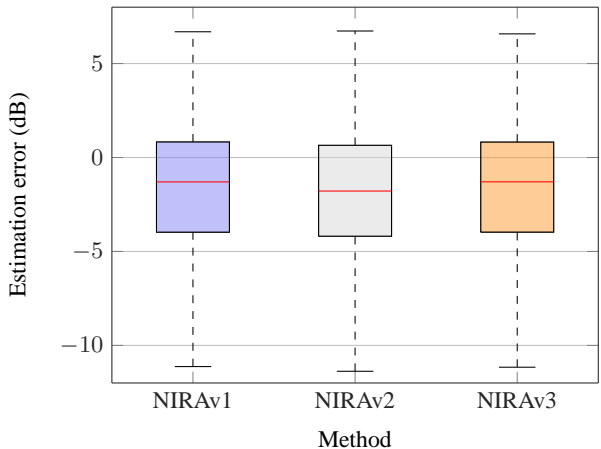
\begin{figure}[htc!]
\centering
%
%
%
%
\begin{tikzpicture}

\begin{axis}[%
width=0.950920245398773\figurewidthx,
height=\figureheightx,
at={(0\figurewidthx,0\figureheightx)},
scale only axis,
separate axis lines,
every outer x axis line/.append style={white!15!black},
every x tick label/.append style={font=\color{white!15!black}},
xmin=0.5,
xmax=3.5,
xtick={1,2,3},
xticklabels={{NIRAv1},{NIRAv2},{NIRAv3}},
xlabel={Method},
every outer y axis line/.append style={white!15!black},
every y tick label/.append style={font=\color{white!15!black}},
ylabel style={yshift=\yyshift},xlabel style={yshift=\xyshift},
ymin=-12,
ymax=8,
ylabel={Estimation error (dB)},
ymajorgrids,
yminorgrids
]

\addplot[area legend,solid,fill=mycolor3,opacity=5.000000e-01,draw=black,forget plot]
table[row sep=crcr] {%
x	y\\
2.775	-3.9733084415\\
2.775	0.8216500675\\
3.225	0.8216500675\\
3.225	-3.9733084415\\
2.775	-3.9733084415\\
}--cycle;

\addplot[area legend,solid,fill=mycolor1,opacity=5.000000e-01,draw=black,forget plot]
table[row sep=crcr] {%
x	y\\
1.775	-4.191206183\\
1.775	0.6484641132\\
2.225	0.6484641132\\
2.225	-4.191206183\\
1.775	-4.191206183\\
}--cycle;

\addplot[area legend,solid,fill=mycolor2,opacity=5.000000e-01,draw=black,forget plot]
table[row sep=crcr] {%
x	y\\
0.775	-3.9765332645\\
0.775	0.82990914\\
1.225	0.82990914\\
1.225	-3.9765332645\\
0.775	-3.9765332645\\
}--cycle;

\addplot [color=black,dashed,forget plot]
  table[row sep=crcr]{%
1	0.82990914\\
1	6.691758815\\
};
\addplot [color=black,dashed,forget plot]
  table[row sep=crcr]{%
2	0.6484641132\\
2	6.732338815\\
};
\addplot [color=black,dashed,forget plot]
  table[row sep=crcr]{%
3	0.8216500675\\
3	6.579828815\\
};
\addplot [color=black,dashed,forget plot]
  table[row sep=crcr]{%
1	-11.1264905\\
1	-3.9765332645\\
};
\addplot [color=black,dashed,forget plot]
  table[row sep=crcr]{%
2	-11.3768605\\
2	-4.191206183\\
};
\addplot [color=black,dashed,forget plot]
  table[row sep=crcr]{%
3	-11.1605405\\
3	-3.9733084415\\
};
\addplot [color=black,solid,forget plot]
  table[row sep=crcr]{%
0.8875	6.691758815\\
1.1125	6.691758815\\
};
\addplot [color=black,solid,forget plot]
  table[row sep=crcr]{%
1.8875	6.732338815\\
2.1125	6.732338815\\
};
\addplot [color=black,solid,forget plot]
  table[row sep=crcr]{%
2.8875	6.579828815\\
3.1125	6.579828815\\
};
\addplot [color=black,solid,forget plot]
  table[row sep=crcr]{%
0.8875	-11.1264905\\
1.1125	-11.1264905\\
};
\addplot [color=black,solid,forget plot]
  table[row sep=crcr]{%
1.8875	-11.3768605\\
2.1125	-11.3768605\\
};
\addplot [color=black,solid,forget plot]
  table[row sep=crcr]{%
2.8875	-11.1605405\\
3.1125	-11.1605405\\
};
\addplot [color=black,solid,forget plot]
  table[row sep=crcr]{%
0.775	-3.9765332645\\
0.775	0.82990914\\
1.225	0.82990914\\
1.225	-3.9765332645\\
0.775	-3.9765332645\\
};
\addplot [color=black,solid,forget plot]
  table[row sep=crcr]{%
1.775	-4.191206183\\
1.775	0.6484641132\\
2.225	0.6484641132\\
2.225	-4.191206183\\
1.775	-4.191206183\\
};
\addplot [color=black,solid,forget plot]
  table[row sep=crcr]{%
2.775	-3.9733084415\\
2.775	0.8216500675\\
3.225	0.8216500675\\
3.225	-3.9733084415\\
2.775	-3.9733084415\\
};
\addplot [color=red,solid,forget plot]
  table[row sep=crcr]{%
0.775	-1.29882586\\
1.225	-1.29882586\\
};
\addplot [color=red,solid,forget plot]
  table[row sep=crcr]{%
1.775	-1.7850162345\\
2.225	-1.7850162345\\
};
\addplot [color=red,solid,forget plot]
  table[row sep=crcr]{%
2.775	-1.2944077345\\
3.225	-1.2944077345\\
};
\end{axis}
\end{tikzpicture}%
	\caption{Distribution of the DRR estimation errors for each method using ACE Challenge evaluation dataset.}
	\label{fig:BoxPlot_DRR_ACE}
\end{figure} 

\begin{figure}[htc!]
\centering
%
%
%
%
\begin{tikzpicture}

\begin{axis}[%
width=0.950920245398773\figurewidthx,
height=\figureheightx,
at={(0\figurewidthx,0\figureheightx)},
scale only axis,
separate axis lines,
every outer x axis line/.append style={white!15!black},
every x tick label/.append style={font=\color{white!15!black}},
xmin=0.5,
xmax=3.5,
xtick={1,2,3},
xticklabels={{NIRAv1},{NIRAv2},{NIRAv3}},
xlabel={Method},
every outer y axis line/.append style={white!15!black},
every y tick label/.append style={font=\color{white!15!black}},
ylabel style={yshift=\yyshift},xlabel style={yshift=\xyshift},
ymin=-100,
ymax=80,
ylabel={Estimation error (\%)},
ymajorgrids,
yminorgrids
]

\addplot[area legend,solid,fill=mycolor3,opacity=1.000000e-01,draw=black,forget plot]
table[row sep=crcr] {%
x	y\\
2.775	-47.139109903365\\
2.775	-3.11121205248605\\
3.225	-3.11121205248605\\
3.225	-47.139109903365\\
2.775	-47.139109903365\\
}--cycle;

\addplot[area legend,solid,fill=mycolor1,opacity=1.000000e-01,draw=black,forget plot]
table[row sep=crcr] {%
x	y\\
1.775	-44.7466850454296\\
1.775	-10.6023495900039\\
2.225	-10.6023495900039\\
2.225	-44.7466850454296\\
1.775	-44.7466850454296\\
}--cycle;

\addplot[area legend,solid,fill=mycolor2,opacity=1.000000e-01,draw=black,forget plot]
table[row sep=crcr] {%
x	y\\
0.775	-45.098620086317\\
0.775	-6.70507433763291\\
1.225	-6.70507433763291\\
1.225	-45.098620086317\\
0.775	-45.098620086317\\
}--cycle;

\addplot [color=black,dashed,forget plot]
  table[row sep=crcr]{%
1	-6.70507433763291\\
1	50.8334520147805\\
};
\addplot [color=black,dashed,forget plot]
  table[row sep=crcr]{%
2	-10.6023495900039\\
2	40.2555903765856\\
};
\addplot [color=black,dashed,forget plot]
  table[row sep=crcr]{%
3	-3.11121205248605\\
3	62.8969436714088\\
};
\addplot [color=black,dashed,forget plot]
  table[row sep=crcr]{%
1	-75.6870505147216\\
1	-45.098620086317\\
};
\addplot [color=black,dashed,forget plot]
  table[row sep=crcr]{%
2	-85.6869847612255\\
2	-44.7466850454296\\
};
\addplot [color=black,dashed,forget plot]
  table[row sep=crcr]{%
3	-77.7980481619978\\
3	-47.139109903365\\
};
\addplot [color=black,solid,forget plot]
  table[row sep=crcr]{%
0.8875	50.8334520147805\\
1.1125	50.8334520147805\\
};
\addplot [color=black,solid,forget plot]
  table[row sep=crcr]{%
1.8875	40.2555903765856\\
2.1125	40.2555903765856\\
};
\addplot [color=black,solid,forget plot]
  table[row sep=crcr]{%
2.8875	62.8969436714088\\
3.1125	62.8969436714088\\
};
\addplot [color=black,solid,forget plot]
  table[row sep=crcr]{%
0.8875	-75.6870505147216\\
1.1125	-75.6870505147216\\
};
\addplot [color=black,solid,forget plot]
  table[row sep=crcr]{%
1.8875	-85.6869847612255\\
2.1125	-85.6869847612255\\
};
\addplot [color=black,solid,forget plot]
  table[row sep=crcr]{%
2.8875	-77.7980481619978\\
3.1125	-77.7980481619978\\
};
\addplot [color=black,solid,forget plot]
  table[row sep=crcr]{%
0.775	-45.098620086317\\
0.775	-6.70507433763291\\
1.225	-6.70507433763291\\
1.225	-45.098620086317\\
0.775	-45.098620086317\\
};
\addplot [color=black,solid,forget plot]
  table[row sep=crcr]{%
1.775	-44.7466850454296\\
1.775	-10.6023495900039\\
2.225	-10.6023495900039\\
2.225	-44.7466850454296\\
1.775	-44.7466850454296\\
};
\addplot [color=black,solid,forget plot]
  table[row sep=crcr]{%
2.775	-47.139109903365\\
2.775	-3.11121205248605\\
3.225	-3.11121205248605\\
3.225	-47.139109903365\\
2.775	-47.139109903365\\
};
\addplot [color=red,solid,forget plot]
  table[row sep=crcr]{%
0.775	-17.9971070286497\\
1.225	-17.9971070286497\\
};
\addplot [color=red,solid,forget plot]
  table[row sep=crcr]{%
1.775	-23.130372206496\\
2.225	-23.130372206496\\
};
\addplot [color=red,solid,forget plot]
  table[row sep=crcr]{%
2.775	-23.8844716150725\\
3.225	-23.8844716150725\\
};
\end{axis}
\end{tikzpicture}%
	\caption{Distribution of the \T60 estimation errors for each method using ACE Challenge evaluation dataset.}
	\label{fig:BoxPlot_T60_ACE}
\end{figure}
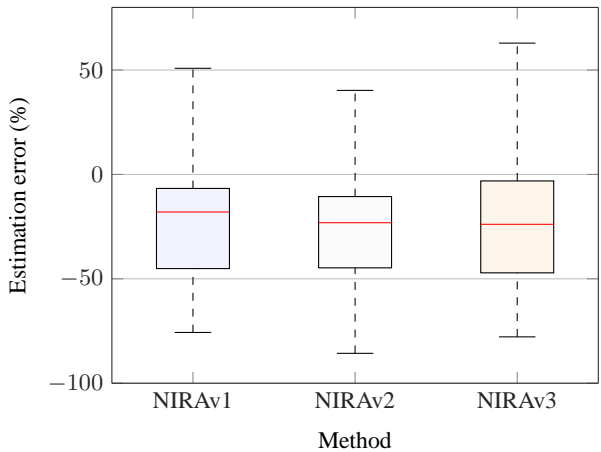

An analysis of the performance of the best approaches to estimate DRR and \T60 is shown in Fig. \ref{fig:BoxPlot_DRR_ACE} and \ref{fig:BoxPlot_T60_ACE} respectively for each noise condition. These figures suggest that babble noise provides the lowest RMSD for DRR estimation whereas fan noise in the recordings brings higher DRR estimation errors. On the contrary, fan noise provides the lowest \T60 deviation and babble noise brings the highest \T60 estimation errors. 

\setlength\figureheightx{4cm}
\begin{figure}[htc!]
\centering
%
%
%
%
\begin{tikzpicture}
\begin{axis}[%
width=\figurewidthx,
height=0.943083028715049\figureheightx,
at={(0\figurewidthx,0\figureheightx)},
area legend,
scale only axis,
separate axis lines,
every outer x axis line/.append style={white!15!black},
every x tick label/.append style={font=\color{white!15!black}},
xmin=0.5,
xmax=3.5,
xtick={1,2,3},
xticklabels={{Low},{Medium},{High}},
xlabel={Noise level},
every outer y axis line/.append style={white!15!black},
every y tick label/.append style={font=\color{white!15!black}},
ylabel style={yshift=\yyshift},xlabel style={yshift=\xyshift},
ymin=0,
ymax=4.5,
ylabel={$\mathrm{RMSD_{DRR}\; (dB)}$},
legend style={at={(0.5,1.03)},anchor=south,legend columns=3,draw=white!15!black,fill=white,legend cell align=left}
]
\addplot[ybar,bar width=0.0592592592592593\figurewidthx,bar shift=-0.0740740740740741\figurewidthx,draw=black,fill=mycolor2] plot table[row sep=crcr] {%
1	3.93741858895404\\
2	3.86533315278966\\
3	3.82035481884613\\
};
\addlegendentry{Ambient};

\addplot[ybar,bar width=0.0592592592592593\figurewidthx,draw=black,fill=mycolor1] plot table[row sep=crcr] {%
1	3.48208724912881\\
2	3.60641442726618\\
3	3.59901026161469\\
};
\addlegendentry{Babble};

\addplot[ybar,bar width=0.0592592592592593\figurewidthx,bar shift=0.0740740740740741\figurewidthx,draw=black,fill=mycolor3] plot table[row sep=crcr] {%
1	4.27826639113938\\
2	4.00846729650127\\
3	3.89106055000361\\
};
\addlegendentry{Fan};

\end{axis}
\end{tikzpicture}%
	\caption{Performance of NIRAv3 estimating DRR on the ACE Challenge evaluation dataset for different noise conditions.}
	\label{fig:NIRAv3_DRR_ACE}
\end{figure}
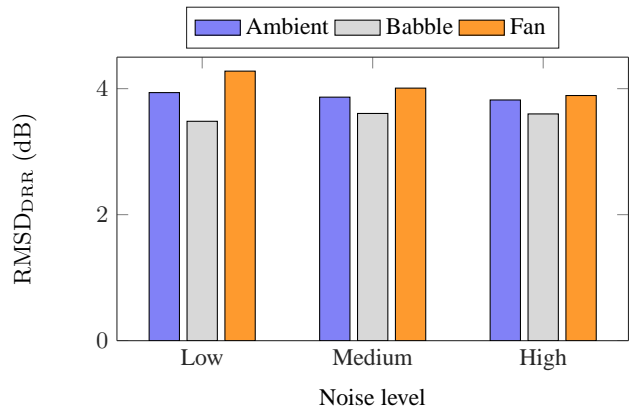 
\begin{figure}[htc!]
\centering
%
%
%
%
\begin{tikzpicture}

\begin{axis}[%
width=\figurewidthx,
height=0.943083028715049\figureheightx,
at={(0\figurewidthx,0\figureheightx)},
area legend,
scale only axis,
separate axis lines,
every outer x axis line/.append style={white!15!black},
every x tick label/.append style={font=\color{white!15!black}},
xmin=0.5,
xmax=3.5,
xtick={1,2,3},
xticklabels={{Low},{Medium},{High}},
xlabel={Noise level},
every outer y axis line/.append style={white!15!black},
every y tick label/.append style={font=\color{white!15!black}},
ylabel style={yshift=\yyshift},xlabel style={yshift=\xyshift},
ymin=0,
ymax=60,
ylabel={$\mathrm{RMSD_{T_{60}}\;(\%)}$},
legend style={at={(0.5,1.03)},anchor=south,legend columns=3,draw=white!15!black,fill=white,legend cell align=left}
]
\addplot[ybar,bar width=0.0592592592592593\figurewidthx,bar shift=-0.0740740740740741\figurewidthx,draw=black,fill=mycolor2] plot table[row sep=crcr] {%
1	42.5683261262086\\
2	39.9051469618086\\
3	38.5652141849418\\
};
\addlegendentry{Ambient};

\addplot[ybar,bar width=0.0592592592592593\figurewidthx,draw=black,fill=mycolor1] plot table[row sep=crcr] {%
1	53.7714656568944\\
2	49.6207161413583\\
3	47.7384042064029\\
};
\addlegendentry{Babble};

\addplot[ybar,bar width=0.0592592592592593\figurewidthx,bar shift=0.0740740740740741\figurewidthx,draw=black,fill=mycolor3] plot table[row sep=crcr] {%
1	37.6793600139504\\
2	37.6738226374897\\
3	37.7924802780423\\
};
\addlegendentry{Fan};

\end{axis}
\end{tikzpicture}%
	\caption{Performance of NIRAv1 estimating \T60 on the ACE Challenge evaluation dataset for different noise conditions.}
	\label{fig:NIRAv1_T60_ACE}
\end{figure}
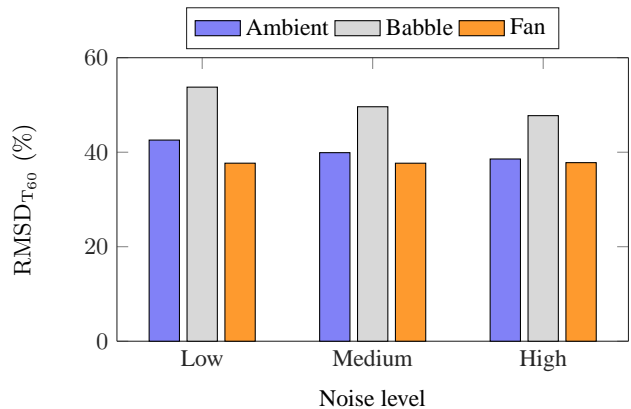

\section{Conclusion}
\label{sec:V}
We have presented in this paper three data-driven approaches to estimate full-band DRR and \T60 from single-channel reverberant speech. These approaches are based on training a BLSTM with different datasets. Additionally, we explored the combination of these networks trained with different datasets by employing a SVR.
The best DRR estimation performance was achieved with NIRAv3, $\mathrm{RMSD_{DRR}}=3.84\; \mathrm{dB}$ with IQR = 4.79~dB and median of -1.3~dB. This is based on training with different databases several BLSTMs and combining their individual time averaged estimations with a SVR. On the other hand, NIRAv1 provides the best \T60 estimation performance, $\mathrm{RMSD_{T_{60}}}= 43.19\; \%$ with IQR = 44~\% and median of -23.88~\%. This configuration is based on training a BLSTM employing only the ACE Challenge development dataset.

Moreover, the performance of these approaches was tested with 10~\% of the ACE Challenge development files, not previously used in the training process, i.e. {\evalset}. The best performance of DRR and \T60 was obtained with NIRAv3 and NIRAv1 respectively, as it occurs on ACE Challenge evaluation dataset. However, the deviations were considerably lower, $\mathrm{RMSD_{DRR}}=0.63\; \mathrm{dB}$ with IQR = 0.7~dB and median of -0.01~dB for DRR estimation and $\mathrm{RMSD_{T_{60}}}= 3.18\; \%$  with IQR = 3.23~\% and median of 0.34~\% for \T60 estimation, which may indicate an overfitting problem in the training process.

\footnotesize
\bibliographystyle{IEEEtran}
\bibliography{References}
%
%
%
%
%
%
%
%
%

\end{sloppy}
\end{document}